\documentclass[runningheads]{llncs}
\usepackage{graphicx}
\begin{document}
\title{Histopathological Stain Transfer using Style Transfer Network with Adversarial Loss} 
\titlerunning{Hist. Stain Transfer using Style Transfer Net. with Adv. Loss}

\author{Harshal Nishar\inst{1} \and
Nikhil Chavanke\inst{1,2} \and
Nitin Singhal\inst{1}}
%
\authorrunning{H. Nishar et al.}
%
\institute{Aira Matrix, Mumbai, India \\
\email{\{harshal.nishar, nitin.singhal\}@airamatrix.com} \\
\url{https://airamatrix.com/}
\and
Indian Institute of Technology, Bombay, India \\
\email{nikhilchavanke21@gmail.com}
}
\maketitle              
\begin{abstract}
Deep learning models that are trained on histopathological images obtained from a single lab and/or scanner give poor inference performance on images obtained from another scanner/lab with a different staining protocol. In recent years, there has been a good amount of research done for image stain normalization to address this issue. In this work, we present a novel approach for the stain normalization problem using fast neural style transfer coupled with adversarial loss. We also propose a novel stain transfer generator network based on High-Resolution Network (HRNet) which requires less training time and gives good generalization with few paired training images of reference stain and test stain. This approach has been tested on Whole Slide Images (WSIs) obtained from 8 different labs, where images from one lab were treated as a reference stain. A deep learning model was trained on this stain and the rest of the images were transferred to it using the corresponding stain transfer generator network. Experimentation suggests that this approach is able to successfully perform stain normalization with good visual quality and provides better inference performance compared to not applying stain normalization. 

\end{abstract}
\section{Introduction}
\begin{figure}[t]
\centering\includegraphics[width=0.85\textwidth]{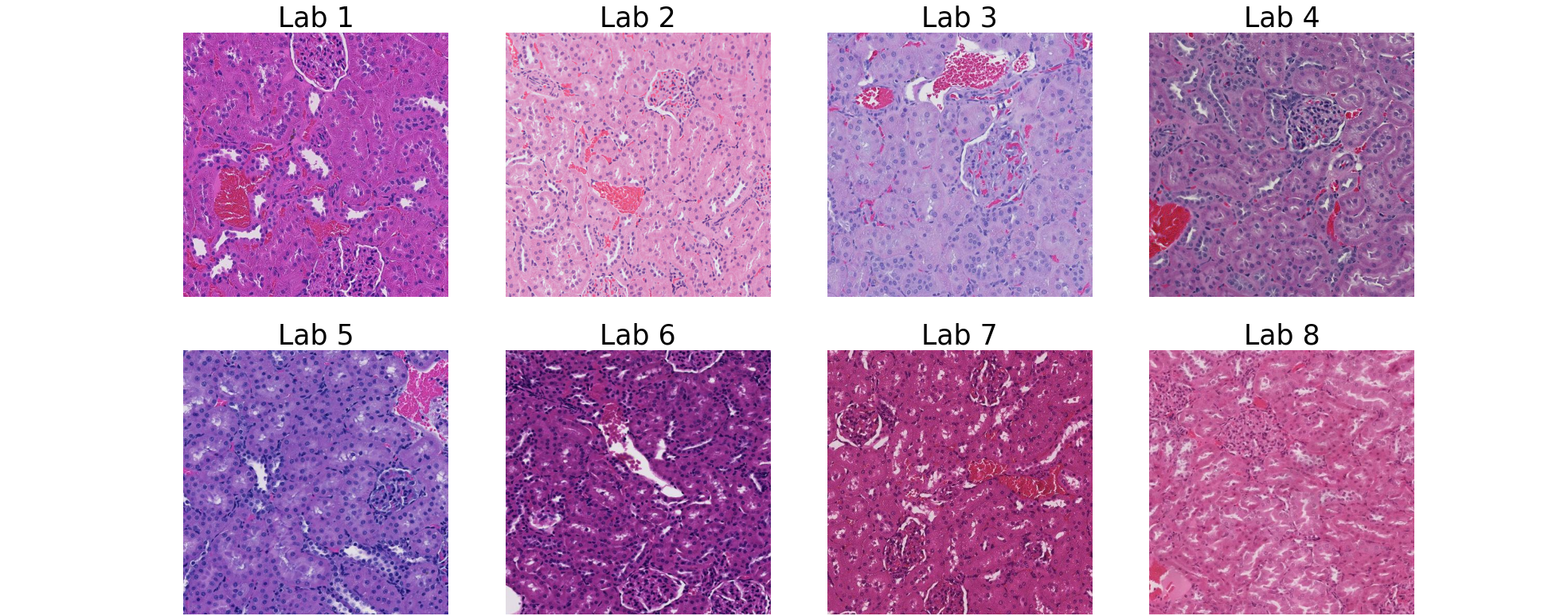}
\caption{Sample patches taken from WSI images of H\&E stained tissues of Wistar rat kidney obtained from 8 different laboratories.}
\label{fig1}
\end{figure}
Histopathological whole slide images (WSI) are generated by scanning tissue images using microscopic scanners. Hematoxylin and Eosin (H\&E) stain is the most commonly used stain for these tissue images. Computer aided histopathological image analysis has gained a significant momentum, and there is an increasing use of supervised deep learning (DL) classification and segmentation models for the same. However, histopathological images obtained from one lab differ significantly from other labs, as every lab has a different staining protocol (Fig.~\ref{fig1}). Same tissue slides that are scanned using different scanners may also generate different looking images due to difference in color calibration. Despite great generalization ability of DL models, stain variation can pose a serious problem for DL models which are generally trained using images sourced from one or few labs. Various image stain normalization techniques have been proposed to solve this stain variation problem. Experimentation has shown that stain normalization can improve the accuracy of DL models~\cite{Ciompi2017TheIO}. 
Recently few approaches have been proposed using deep generative neural networks for image stain normalization which have shown promising results~\cite{Zanjani2018DCGMM},~\cite{BentaiebAdvStainTx},~\cite{StainGAN}. Building onto that, we present an image to image transformation network, called stain transfer generator network, based on High-Resolution Network (HRNet)~\cite{sun2019deep} architecture to transform images from one stain to another.
\subsubsection{Contributions}
1. We propose to use neural style transfer with addition of adversarial loss for image stain	normalization. To the best of our knowledge, this is the first work which couples neural style transfer with GAN based framework for histopathological stain normalization
2. We suggest a simple yet effective modification to HRNet: a direct skip connection from input to output. This helps in faster training and improved convergence of stain transfer generator network.
3. We have tested the proposed method on data obtained from 8 different labs.
\section{Existing Approaches for Stain Normalization}
Histogram matching of RGB channnels of input image with corresponding channel of reference image is the simplest approach, but it requires similar distribution of various tissue components. Reinhard et al.~\cite{Reinhard} uses global image mean and standard deviation in LAB color space for image normalization. Such global methods do not consider local variation in color distribution and hence lead to inaccurate color mapping. If applied on local image patches then there is no consistency in transformed patches across the WSI. Bejnordi~\cite{Bejnordi2016StainSS} uses color and shape based pixel classification into dye types and matches class specific mean and standard deviation.

Color deconvolution~\cite{ruifrok2001quantification} based stain separation methods have been proposed, which separate images into H and E stain components and apply normalization on each of them separately. A pioneering work in automatic color deconvolution matrix computation uses non-negative matrix factorization for stain unmixing~\cite{NIPS2003_NMF}.
Recent work by Vahadane et. al~\cite{Vahadane2016SPCN} uses sparse non-negative matrix factorization (SNMF) for this task. Any stain separation method transforms three channel color image into image having two dimensional basis with each basis vector representing H and E color component.
In all the reconstructed images, we have observed red color of RBCs getting wrongly transformed to pink color of cytoplasm. Loss of such vital color information can cause significant degradation of DL models which depend on RBC features. Also, this method does not classify whether the component belongs to H stain or E stain automatically.

Recently, some authors have used deep learning to solve stain color normalization problem. In~\cite{StaNoSA}, image patch is divided into several classes using K-means clustering on features obtained from trained auto-encoders and histogram matching is done on each cluster to match image stain. Zanjani et al.~\cite{Zanjani2018DCGMM}, uses deep convolutional GMM to do image clustering and then applies cluster-wise color transfer.  Performance of cluster based color matching techniques depends heavily on clustering accuracy and even slight misclassification of cluster can introduce severe degradation. Some approaches use grayscale image colorization using GAN for stain normalization~\cite{cho2017neural}~\cite{Zanjani2018StainNO},~\cite{Yuan2018}. If we could obtain complete color information from grayscale images, then it would have been easier to train deep learning models directly on the grayscale images and save significant computation time and the need for stain normalization. However, segmentation of some features invariably need color images and hence grayscale image colorization cannot be helpful there. In~\cite{BentaiebAdvStainTx}, authors have jointly trained GAN based stain transformation network along with classifier. CycleGANs have also been used for image stain normalization~\cite{StainGAN}, additionally authors in~\cite{Gadermayr2018WhichWR},~\cite{pmlr-v102-de-bel19a} proposed to use identity loss in addition to cycle consistency loss. In \cite{sun2019deep}, authors have used deep convolutional features for pairing reference image patches with input image patch and then compute a global color transformation matrix~.
\section{Proposed Method}
The problem of histopathological image stain normalization can be formulated as follows:

Let $R$ corresponds to a set of reference stain images, $P$ be a set of input stain images and $T$ be a set of transformed images.
We want to find a transformation $\mathcal{G}: P \rightarrow T$ such that $\mathcal{S}(T,R)$ is minimum and $\mathcal{C}(\mathcal{G}(p), p)$ is minimum $\forall$ $p \in P$ 

\noindent Here $\mathcal{S}(.)$ is stain style similarity measure which defines the similarity between two sets of stains and $\mathcal{C}(.)$ is content similarity measure between input image and transformed image.

This is a highly ill-posed problem where a concrete similarity measure is also not available. Our approach to solve this problem is inspired from neural style transfer~\cite{neuralstyletx} where a pre-trained neural network is used as a feature extractor for style as well as content features. The original neural style transfer method follows iterative optimization approach to transfer each of the input images to desired style on the fly, making it very slow. Instead, here a stain transfer generator CNN is trained using a small set of paired input and reference stain images. Training a neural network to do this task is similar to fast neural style transfer proposed by Johnson et al.~\cite{Johnson2016Perceptual}.
\subsection{Loss Definition}
Let $p$ be input image, $r$ be reference image and $t$ be transformed image (the output of stain transfer generator network), then content and style loss is computed as follows:
\subsubsection{Content Loss} is computed from the content features($F_{ij}^{l}$) obtained by passing the input image $p$ and the transformed image $r$ through a pre-trained CNN (eg. VGG16 or VGG19). Here $F_{ij}^{l}$ corresponds to the features from the $i^{th}$ feature map of the $l^{th}$ layer of the CNN at the $j^{th}$ location.
\begin{equation}
\mathcal{L}_{content}^{l}(p, t)=\frac{1}{2}\sum_{ij} (F_{ij}^{l}(p)-F_{ij}^{l}(t))^{2}
\end{equation}
\subsubsection{Style Loss} is computed from the style representation ($G^{l}$) obtained by passing the reference image $r$ and the transformed image $t$ through a pre-trained CNN (eg. VGG16 or VGG19). Style representations are the Gram matrices of CNN features at each layers. Each element of a Gram matrix at location $ik$ is given by the inner product of the vectorized $i^{th}$ and $k^{th}$ feature maps:
\begin{equation}
G^{l}_{ik} = \sum_{j}F^{l}_{ij} F^{l}_{kj}
\end{equation}
Style loss corresponding to each layer $l$ is given below where $N_l$ is total number of feature maps and $M_l$ is the size of the feature map at layer $l$:
\begin{equation}
\mathcal{E}^{l}(r, t) = \frac{1}{N_l M_l} \sum_{ik}(G^{l}_{ik}(r) - G^{l}_{ik}(t))^2
\end{equation}
Total style loss is computed as the weighted sum of the style loss for each layer
\begin{equation}
\mathcal{L}_{style}(r, t) = \sum_{l}\omega_l\mathcal{E}^l(r, t)
\end{equation}

Further, we propose to add an adversarial loss to the above loss, thus making our network a generative adversarial network. Generative adversarial network first proposed by Goodfellow et el.~\cite{NIPS2014_GAN} has been extensively used in image to image translation tasks due to its ability to generate new images appearing as if they have been drawn from the reference domain.

\begin{figure}
\includegraphics[width=\textwidth]{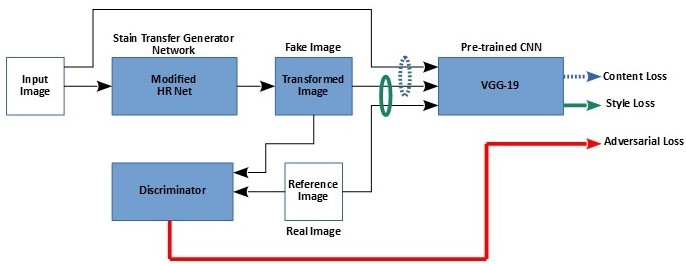}
\caption{Block diagram showing how each of the training losses are computed.}
\label{fig2}
\end{figure}
To compute adversarial loss an additional network called discriminator ($\mathcal{D}$) has to be trained such that it distinguishes between original reference stain images and fake images generated using stain transfer generator network.
\subsubsection{Discriminator Loss} is computed as follows:
\begin{equation}
\mathcal{L}_{dis} = log(1 - \mathcal{D}(r)) + log(\mathcal{D}(t))
\end{equation}
\subsubsection{Adversarial Loss:}
The generator has to generate images that are similar to reference stain images so that the discriminator can not distinguish them from the real samples drawn from reference stain images. For this purpose, we need to minimize the adversarial loss defined as follows:
\begin{equation}
\mathcal{L}_{adv} = log(1 - \mathcal{D}(t))
\end{equation}
\subsubsection{Total Generator Loss} is computed as a weighted combination of content loss, style loss and adversarial loss, where each of the weighting factors ($\lambda_{a}$, $\lambda_{c}$ and $\lambda_{s}$) are model hyper-parameters.
\begin{equation}
\mathcal{L}_{gen} =  \lambda_{a} \mathcal{L}_{adv} +  \lambda_{c} \mathcal{L}_{content} + \lambda_{s} \mathcal{L}_{style}
\end{equation}
To train the stain transfer generator network, the discriminator loss ($\mathcal{L}_{dis}$) and the total generator loss ($\mathcal{L}_{gen}$) are iteratively minimized. While training the discriminator, the generator weights are freezed and vice versa while training the generator, the discriminator weights are freezed. However, the weights of the pre-trained CNN network, which is used only as a deep feature extractor, must be kept constant throughout the training process. Fig.~\ref{fig2} is a block diagram of the complete system that illustrates how various generator losses are computed.
\subsection{Generator Network Architecture}
\begin{figure}
\includegraphics[width=\textwidth]{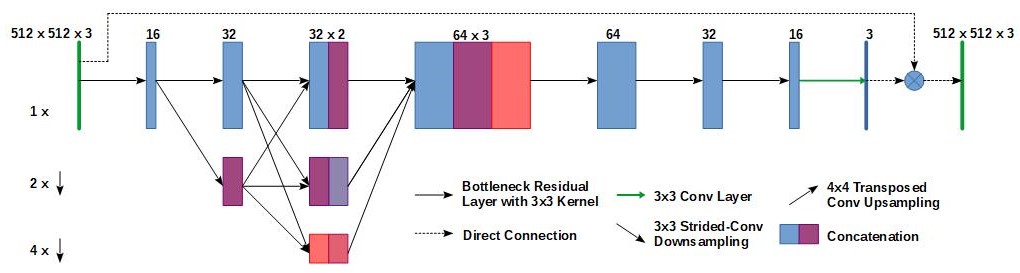}
\caption{Stain Transfer Generator Network Architechure (Modified HRNet).}
\label{fig3}
\end{figure}
In the artistic style transfer, retention of the minute details in the transformed image is not very important. But the stain transfer generator network must retain the high resolution content features in the transformed image. Hence instead of ResNet based style transfer network proposed by Johnson et al, our stain transfer generator architecture is based on High-Resolution Net (HRNet). HRNet has been shown to retain rich high resolution representations, which helps in maintaining the quality of histopathological images. As a major modification to HRNet, a direct skip connection from input to output has been added. Fig.~\ref{fig3} shows full generator network architecture with feature map size at each layers. Let $p$ be the input image and $\mathcal{G}$ be the stain transfer generator network then,
\begin{equation}
\mathcal{G}(p) = p + \mathcal{G}^{\prime}(p)
\end{equation}
Intuition behind the skip connection is that, in the worst case scenario, the stain transfer generator network should at least return the original content image. Due to a direct skip connection, the network has to learn only the residue transformation instead of the actual transformation. This residual learning is much faster and gives better convergence. The proposed stain transfer generator network learns good transformation with as little as 20 paired training images and 50 epochs of training.
\subsection{Implementation Details}
WSIs are divided into non-overlapping patches of 512 x 512 pixels at 10x magnification. Around 20 patches from each input stain is paired to reference stain patch based on content similarity. In stain transfer generator network, all convolutional layers except last one are followed by batch-normalization and ReLU activation. The discriminator follows architectural guidelines given in DCGAN~\cite{radford2015unsupervised} and uses instance-normalization instead of batch-normalization. Also, our discriminator implementation is 16 x 16 PatchGAN~\cite{Isola2016ImagetoImageTW} and hence it classifies 16x16 patch of an image as real or fake. The generator learning rate is kept at 1e-3 and the discriminator learning rate is kept at 1e-5. For training, we have used Adam optimizer and batch size of 4. Content features are taken from second convolutional layer of VGG19 (conv2\_2) pre-trained on ImageNet dataset. For style loss, weight of each layer is kept constant 1.
\section{Experimentation}
\begin{figure}[t]
\includegraphics[width=0.9\textwidth]{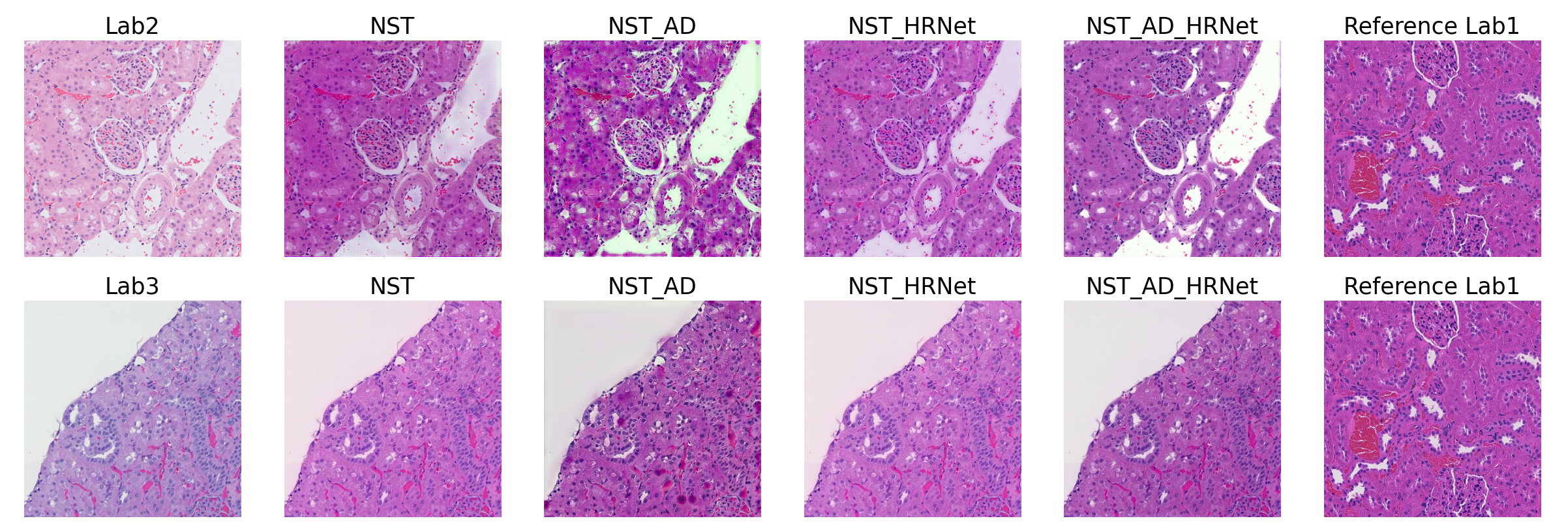}
\caption{This figure shows output of style transfer network with different configuration. Original 512x512 image patches from lab 2 and lab 3 are shown at leftmost side.}
\label{fig4}
\end{figure}

Throughout the experimentation, lab 1 (see Fig. \ref{fig1}) is selected as a reference stain and other images are transformed to this stain. To check the effectiveness of our proposed method, we have compared the stain transfer performance in 4 different settings: 1. neural style transfer using generator network proposed in \cite{Johnson2016Perceptual}(NST) 2. neural style transfer with addition of adversarial loss (NST\_AD) 3. neural style transfer using modified HRNet (NST\_HRNet) and 4. neural style transfer with adversarial loss and modified HRNet (NST\_AD\_HRNet). All the models are trained with their best hyper-parameters. It has been observed that NST gave good results with Lab 3 but very poor results with Lab 2. NST\_AD has generated images looking like reference stain but has distortions. NST\_HRNet has good results with both lab 2 and lab 3 with minute content details preserved. NST\_AD\_HRNet has generated best results for both the labs, with images looking very close to the reference stain (Fig. \ref{fig4}). 

\begin{figure}
\includegraphics[width=0.9\textwidth]{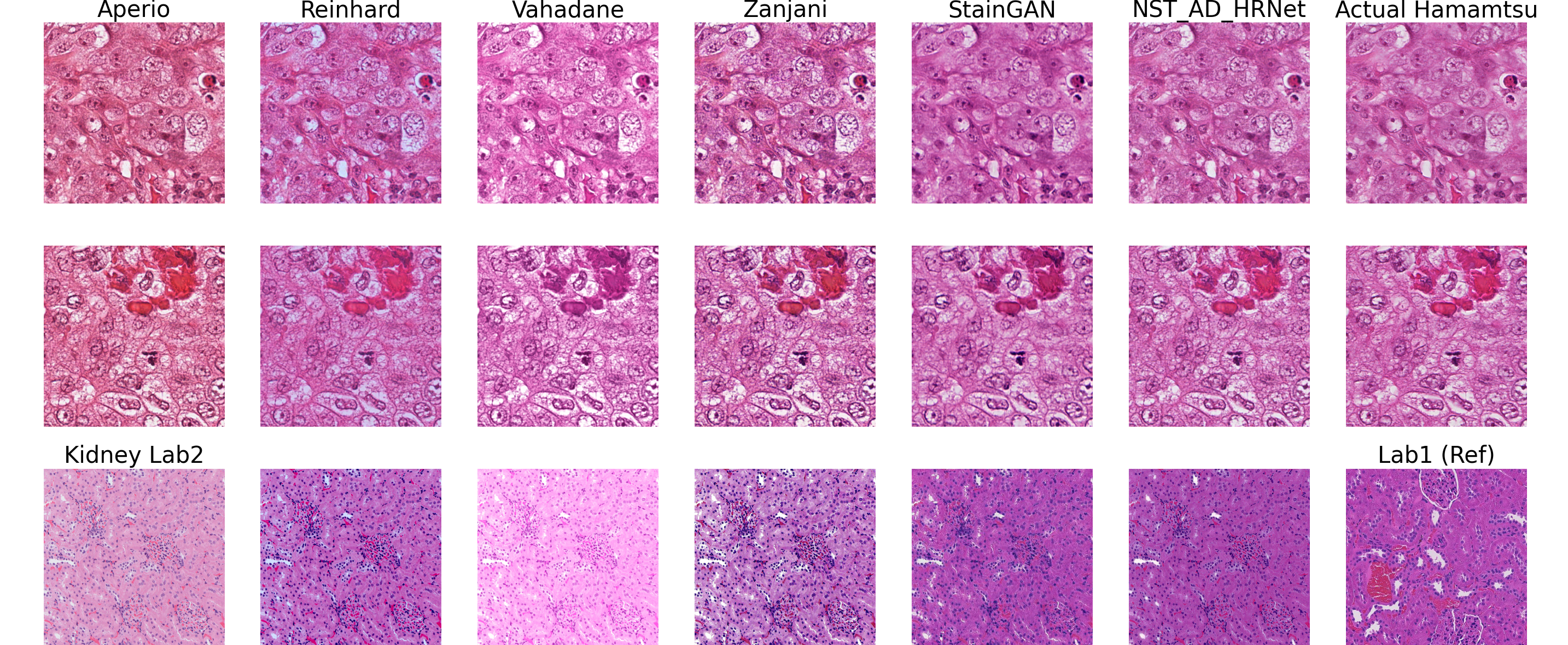}
\caption{Comparison of the proposed approach with the other approaches and with the reference. First two rows show results for mitosis dataset and last row shows results for lab 2 from our kidney dataset.}
\label{fig5}
\end{figure}

For result comparison with other well-known approaches (Reinhard~\cite{Reinhard}, Vahadane~\cite{Vahadane2016SPCN}, Zanjani~\cite{Zanjani2018DCGMM} and StainGAN~\cite{StainGAN}), we use publicly available ICPR2014 Mitosis dataset\footnote{https://mitos-atypia-14.grand-challenge.org/Dataset/} in which the same tissue slides are scanned using two different scanners (Aperio and Hamamatsu). Hamamatsu scanner is taken as reference and Aperio scanner images are transformed to it. In Fig.~\ref{fig5}, we can see that Vahadane's method makes red RBCs pinkish as we had discussed previously. Zanjani's DCGMM produces good output but has sharpened texture and brighter colors as compared to actual reference images. Also, due to large textural variations in kidney images, DCGMM does not produce good clusters and gives poor stain normalized output. StainGAN produces decent output for mitosis images, however, for kidney images it produces images that look like reference stain but with significant textural distortions. We have observed that some of the image similarity measures like SSIM and PSNR are quite misleading and do not indicate true perceptual similarity between images, hence we have used deep features based perceptual similarity measure by Zhang et al.\cite{zhang2018perceptual}, which has been shown to perform significantly better perceptually than other measures. On this measure, the performance of the proposed approach is only 3\% poorer than StainGAN and 20\% better than Zanjani's approach (Tabel~\ref{tab1}). While CycleGAN based StainGAN took more than 72 hours to train, the proposed network got trained in less than 2 hours. Also, our generator network did not suffer from training instability generally observed with GAN. See appendix for the comparison on kidney data from all the labs.
\begin{table}
    \caption{Comparison with deep features based perceptual similarity distance measure (lower is better)}
    \label{tab1}
    \small
    \setlength{\tabcolsep}{3pt}
    \begin{center}
    \begin{tabular}{ c | c       | c        | c       | c        | c}
        Method        & Reinhard & Vahadane & Zanjani & StainGAN & NST\_AD\_HRNet \\
        \hline 
        Mean Distance & 0.325    & 0.205    & 0.191   & 0.148    & 0.153          \\
        Std. Div.     & 0.129    & 0.156    & 0.080   & 0.111    & 0.107          \\
    \end{tabular}
    \end{center}
\end{table}
\begin{table}
\caption{Dice score for the DL model with and without proposed stain normalization.}
\label{tab2}
\setlength{\tabcolsep}{3pt}
\begin{center}
\begin{tabular}{ c | c | c | c | c | c | c | c}
Lab        & Lab 2 & Lab 3 & Lab 4 & Lab 5 & Lab 6 & Lab 7 & Lab 8  \\
\hline 
Without Normalization Dice & 0.183 & 0.754 & 0.804 & 0.718 & 0.805 & 0.593 & 0.641  \\
With Normalization Dice    & 0.437 & 0.904 & 0.815 & 0.900 & 0.899 & 0.787 & 0.771  \\
\end{tabular}
\end{center}
\end{table}

To validate the effectiveness of the proposed approach in improving the accuracy of a pre-existing DL model on new unseen stain images, testing has been done on 7 different stains. We have trained a ResNet50-FCN based Glomeruli segmentation network on image patches taken from lab 1. Color based data augmentations were applied on training data. For lab 2 to lab 8, we trained corresponding stain transfer generator network to transform each of them to lab 1. Table~\ref{tab2} shows the performance of segmentation model with and without proposed stain normalization. There is a significant improvement in model performance due to the proposed stain normalization for all the labs.
\section{Conclusion}
In this paper we presented a novel, fast and effective stain color normalization technique for H\&E stained histopathological images. The proposed method was compared with other well known techniques and gave superior stain normalization performance. Due to a direct skip connection from input to output, stain transfer generator network can be trained quickly and with very few paired training images. This helps in quick adaptation of existing deep learning models for a differently stained data. The proposed method was tested on data obtained from 8 different labs and was able to transform each of them to the reference stain without any distortions in the original content. We believe this work can be extended to staining dyes other than H\&E and can also be used for domain adaptation like transforming images from H\&E to PAP stain.
%
%
%
%
%
\bibliographystyle{splncs04}
\bibliography{references}
\appendix
\section*{Appendix}
\begin{figure}[!htbp]
\includegraphics[width=\textwidth]{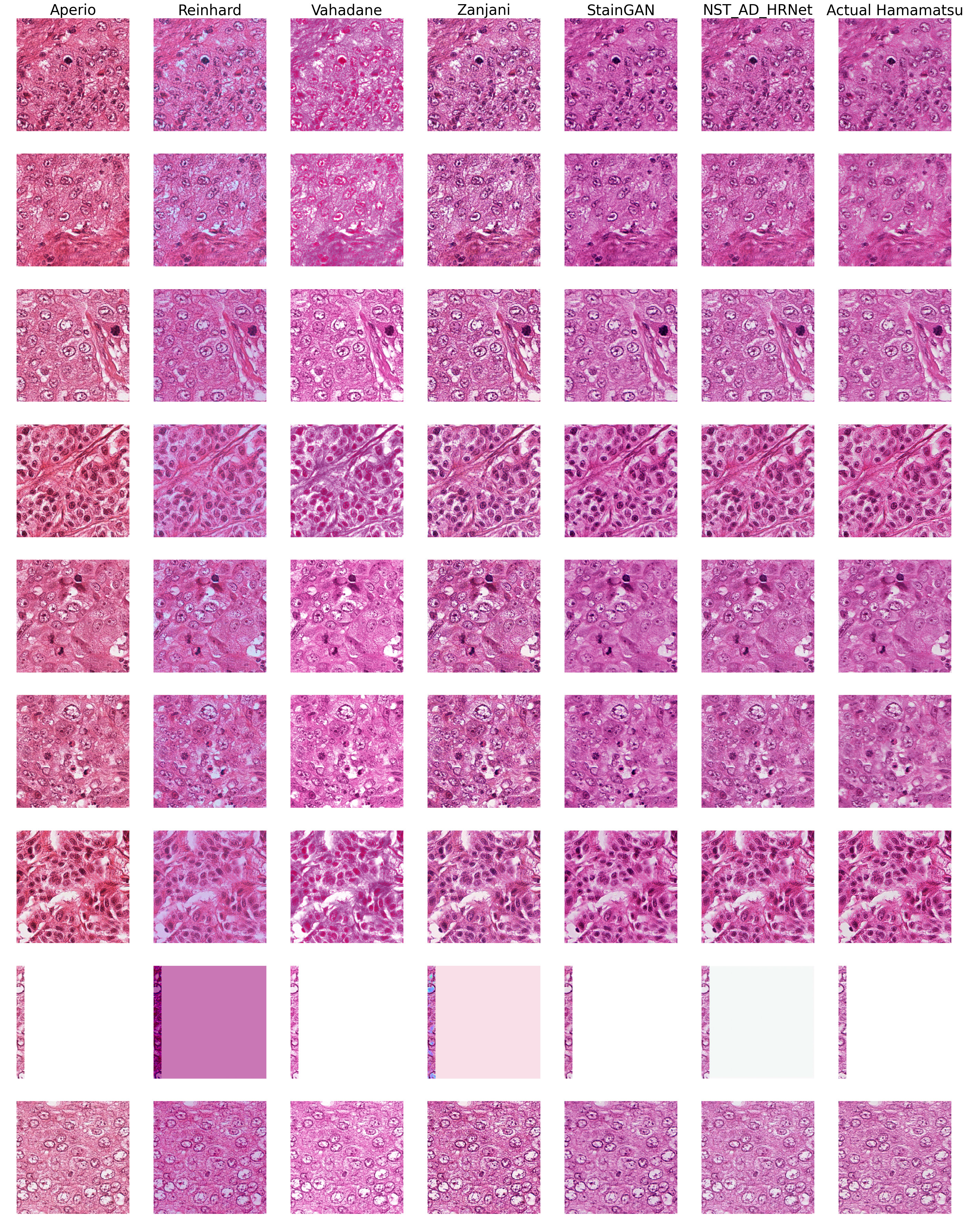}
\caption{Comparison of the proposed approach with the other approaches for stain normalization on the Mitosis data scanned from Aperio scanner to hamamatsu scanner. These image tiles of size 512 x 512 are sampled randomly}
\label{sfig1}
\end{figure}
\begin{figure}
\includegraphics[width=\textwidth]{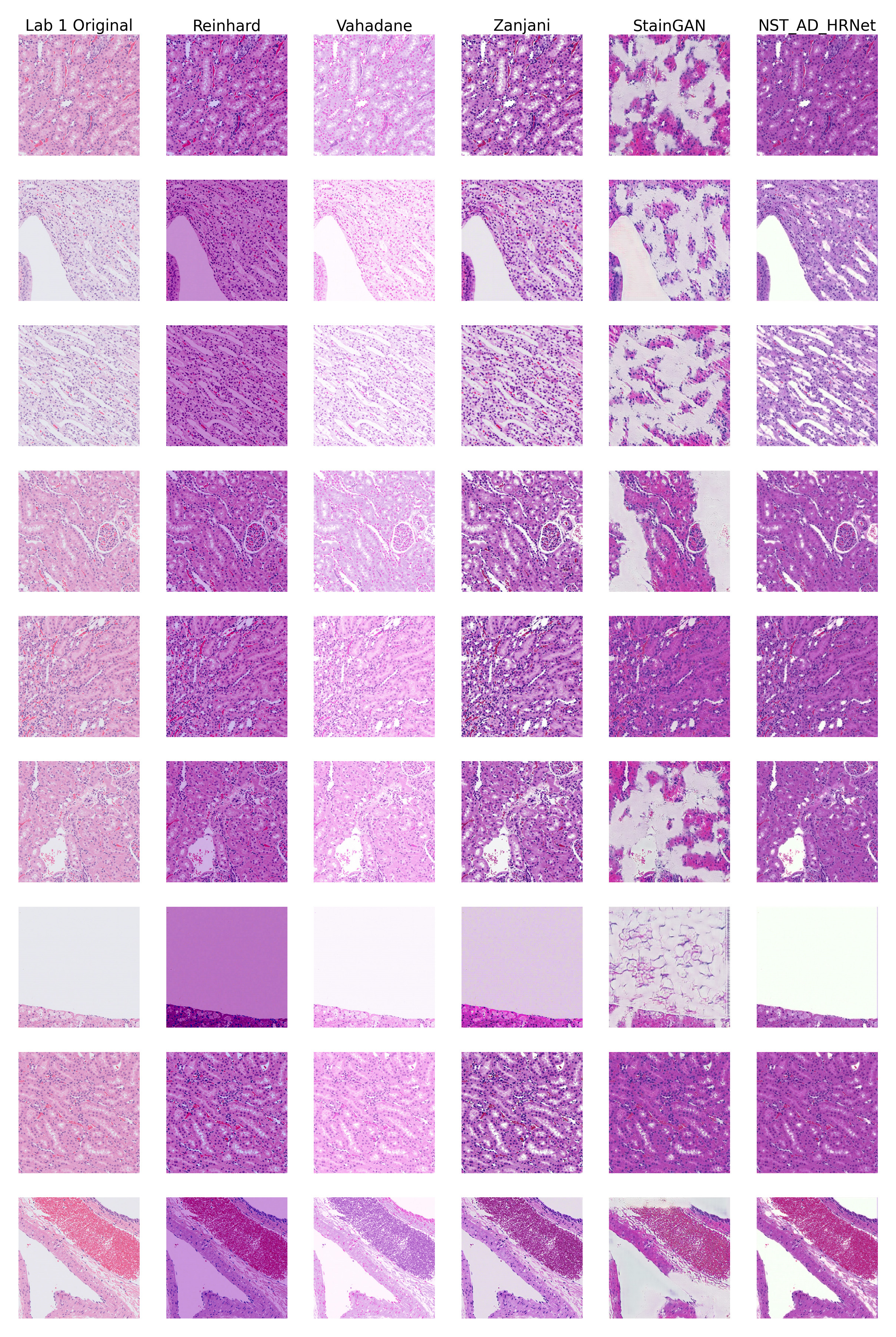}
\caption{Comparison of the proposed approach with the other approaches for stain normalization on the kidney data obtained from lab 2 to reference lab 1. These image tiles of size 512 x 512 are sampled randomly}
\label{sfig2}
\end{figure}
\begin{figure}
\includegraphics[width=\textwidth]{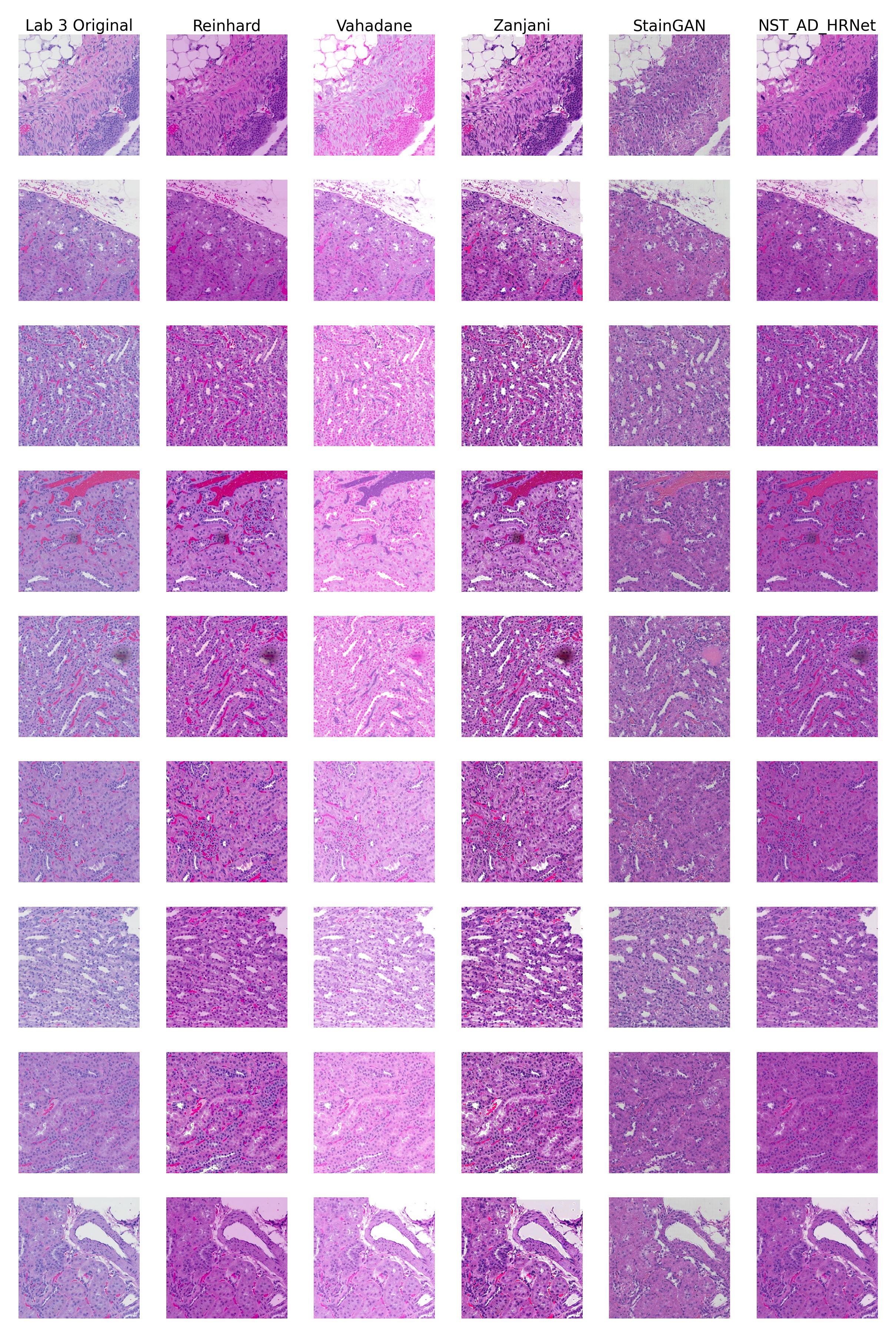}
\caption{Comparison of the proposed approach with the other approaches for stain normalization on the kidney data obtained from lab 3 to reference lab 1. These image tiles of size 512 x 512 are sampled randomly}
\label{sfig3}
\end{figure}
\begin{figure}
\includegraphics[width=\textwidth]{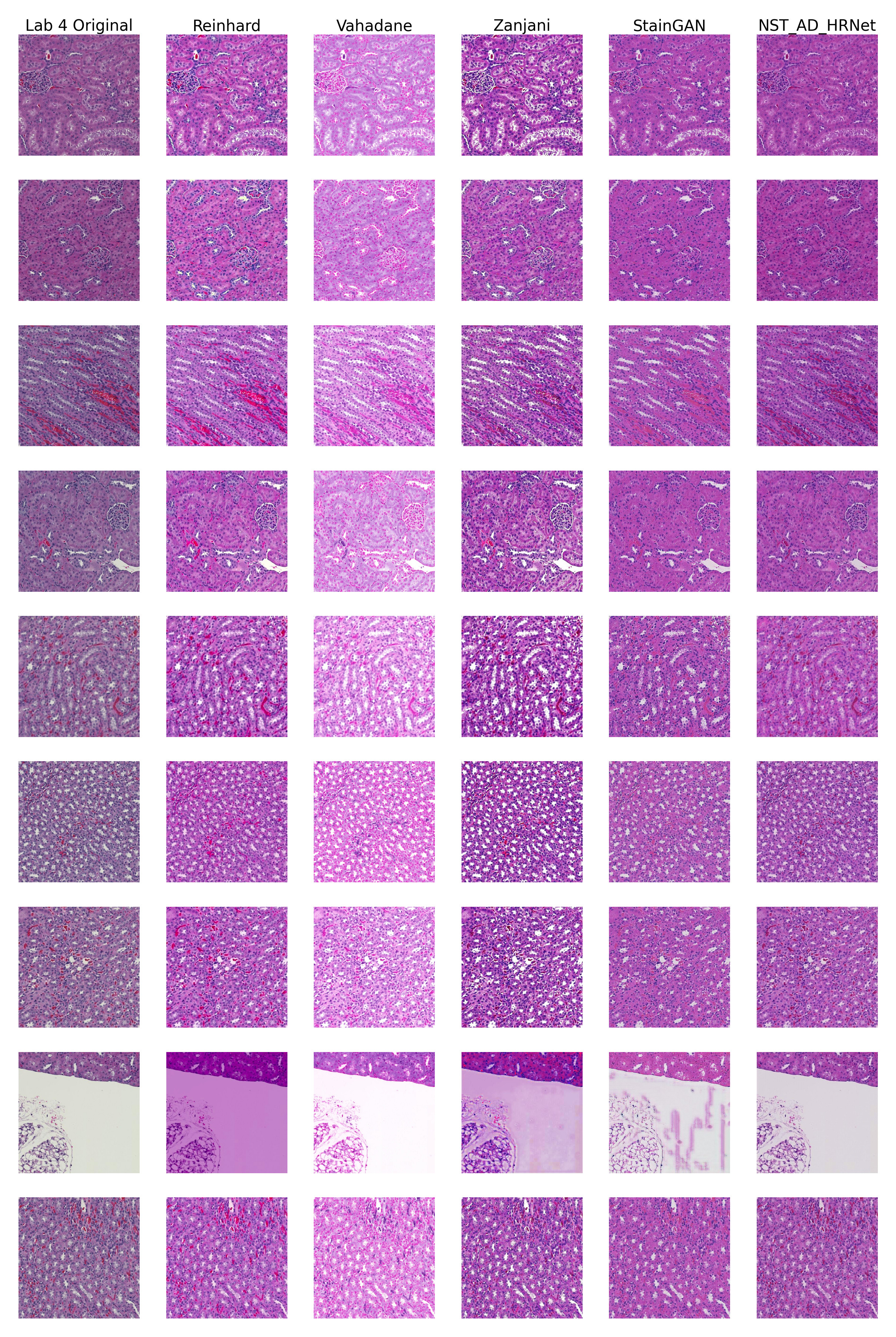}
\caption{Comparison of the proposed approach with the other approaches for stain normalization on the kidney data obtained from lab 4 to reference lab 1. These image tiles of size 512 x 512 are sampled randomly}
\label{sfig4}
\end{figure}
\begin{figure}
\includegraphics[width=\textwidth]{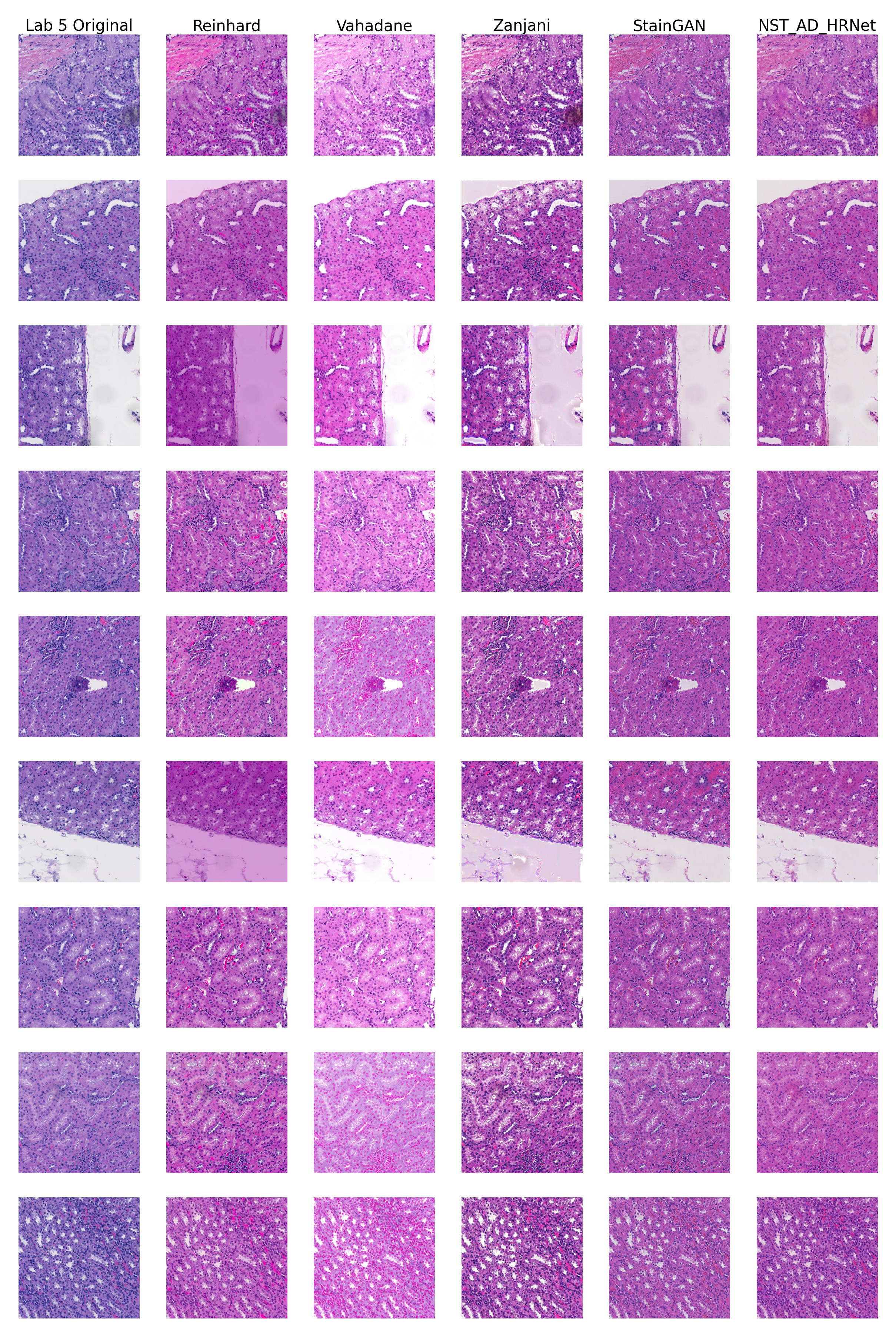}
\caption{Comparison of the proposed approach with the other approaches for stain normalization on the kidney data obtained from lab 5 to reference lab 1. These image tiles of size 512 x 512 are sampled randomly}
\label{sfig5}
\end{figure}
\begin{figure}
\includegraphics[width=\textwidth]{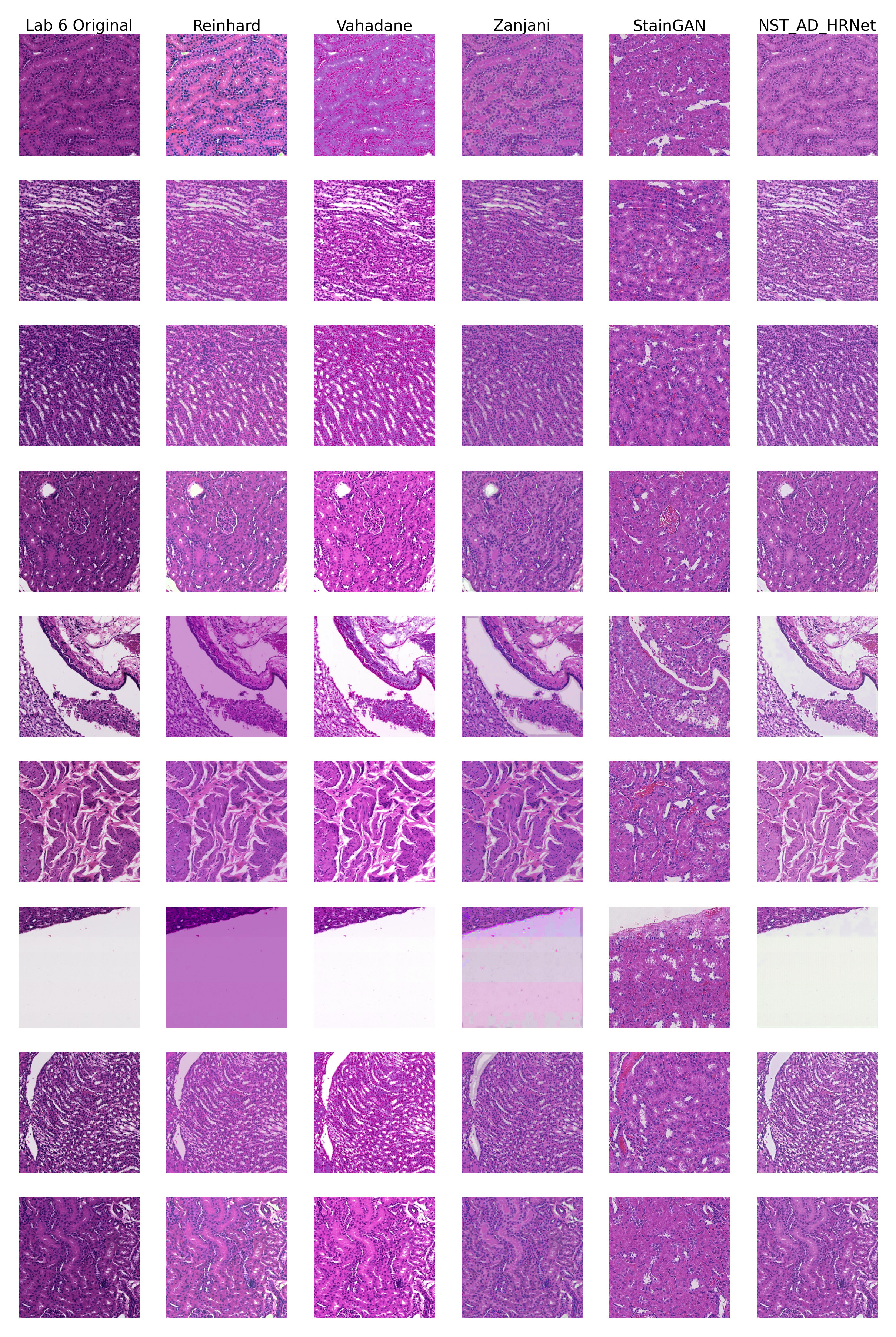}
\caption{Comparison of the proposed approach with the other approaches for stain normalization on the kidney data obtained from lab 6 to reference lab 1. These image tiles of size 512 x 512 are sampled randomly}
\label{sfig6}
\end{figure}
\begin{figure}
\includegraphics[width=\textwidth]{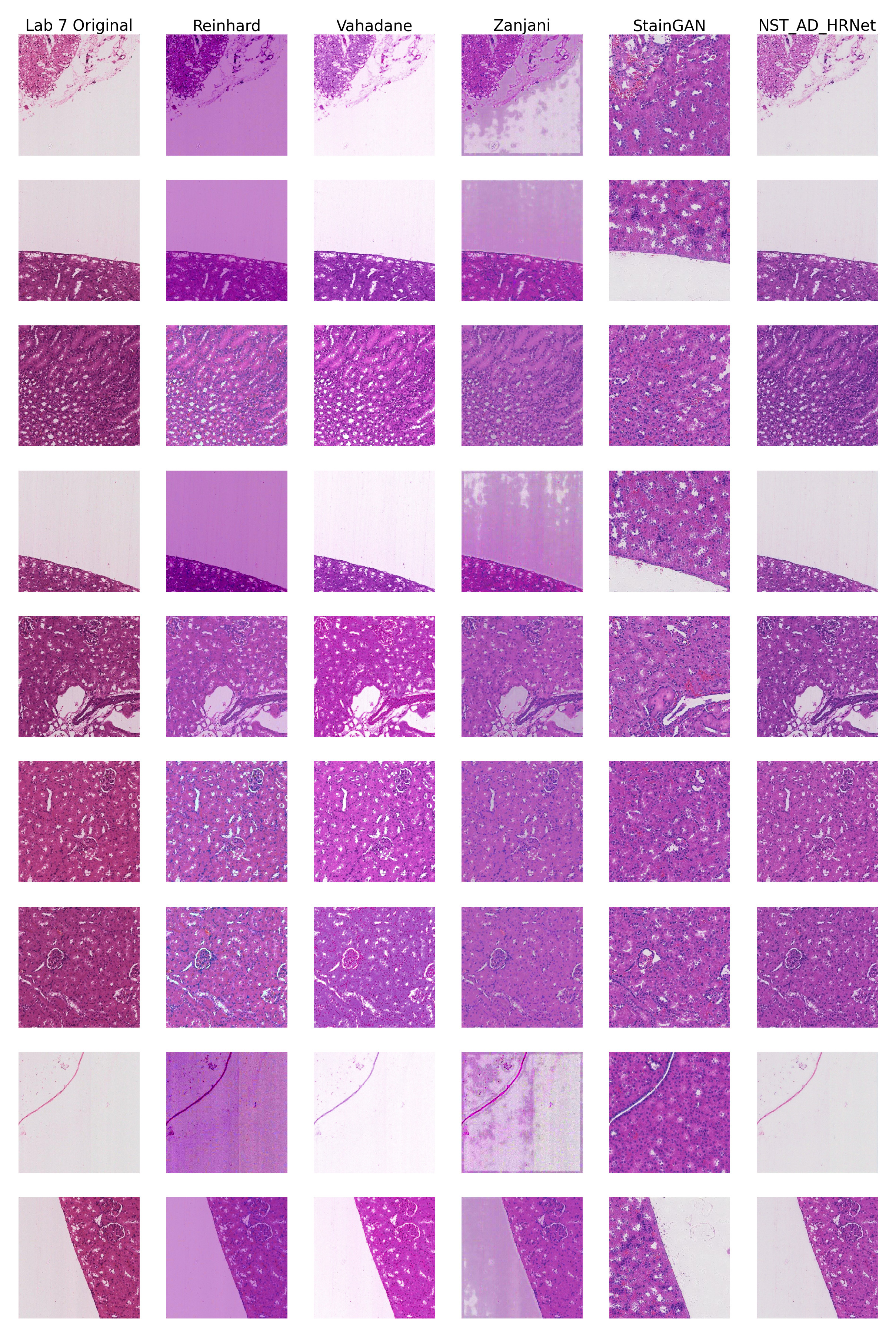}
\caption{Comparison of the proposed approach with the other approaches for stain normalization on the kidney data obtained from lab 7 to reference lab 1. These image tiles of size 512 x 512 are sampled randomly}
\label{sfig7}
\end{figure}
\begin{figure}
\includegraphics[width=\textwidth]{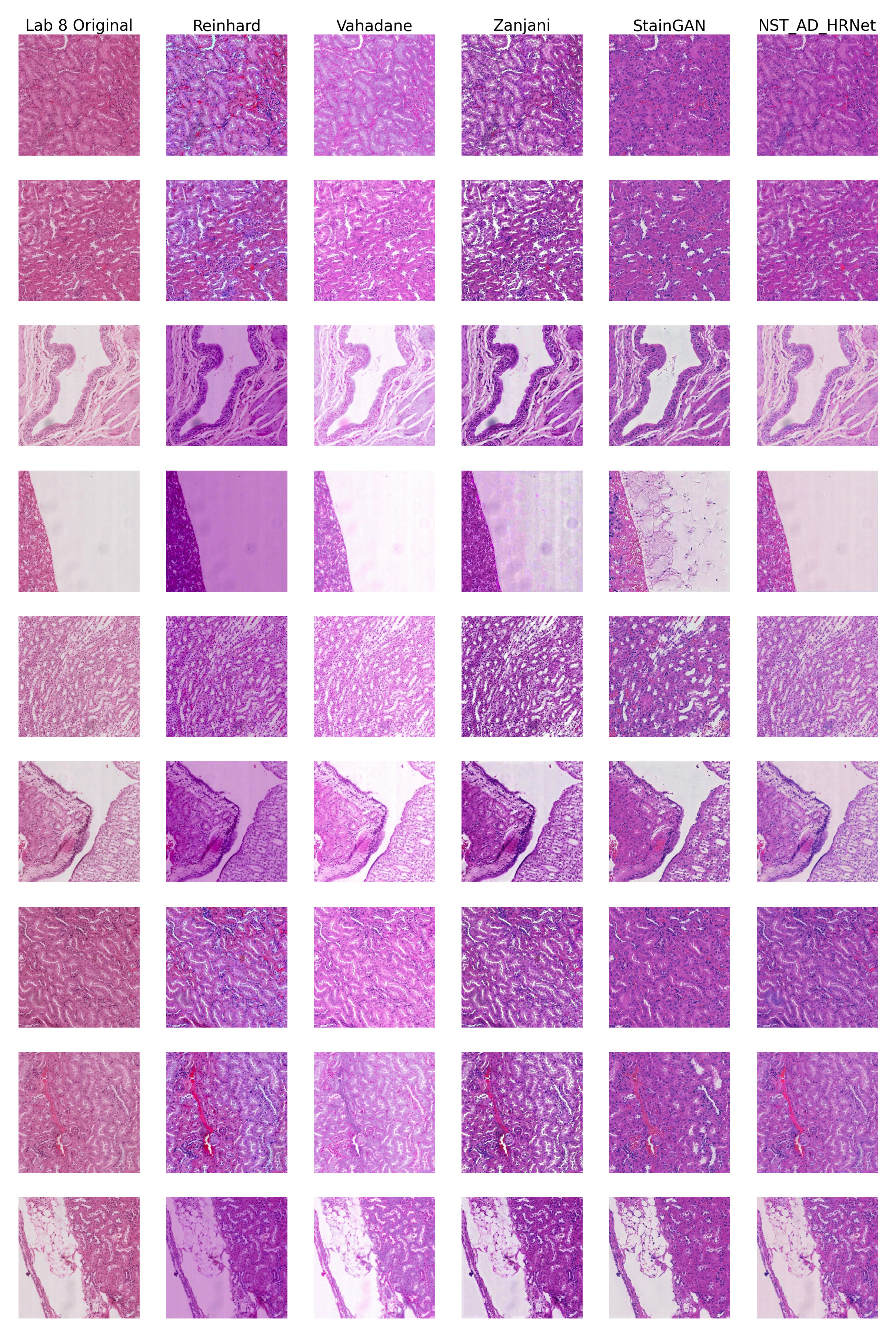}
\caption{Comparison of the proposed approach with the other approaches for stain normalization on the kidney data obtained from lab 8 to reference lab 1. These image tiles of size 512 x 512 are sampled randomly}
\label{sfig8}
\end{figure}

\end{document}